\documentclass[aps, prb, twocolumn, superscriptaddress, floatfix, longbibliography, citeautoscript]{revtex4-2}

\usepackage{graphicx, tabularx}
\usepackage{bm}
\usepackage{amssymb, amsmath}

\usepackage{xcolor}
\definecolor{mblue}{RGB}{0, 118, 186}

\definecolor{mgreen}{RGB}{29, 177, 0}

\definecolor{mred}{RGB}{209, 28, 36}

\usepackage{hyperref}
\hypersetup{
    colorlinks=true,
    linkcolor=blue,     
    urlcolor=blue,
    citecolor=blue
}
\hypersetup{
  pdfstartview={XYZ null null 1.00}, % Sets zoom to 100%
  pdfpagemode=UseNone, % Closes the bookmarks/sidebar panel
  pdfpagelayout=OneColumn % Sets layout to Single Page Continuous
}

\newif\ifptitle
\newif\ifpnumber
\newcounter{para}
\newcommand\ptitle[1]{\par\refstepcounter{para}
{\ifpnumber{\noindent\textcolor{darkgray}{\textbf{\thepara}}\indent}\fi}
{\ifptitle{\textbf{[{#1}]}}\fi}}

\ptitlefalse  % paragraph titles
\pnumberfalse % paragraph numbers

\begin{document}

\title{Rapidly prototyping kagome flat band physics with acoustic metamaterials}

\newcommand{\hphys}{Department of Physics, Harvard University, Cambridge, MA 02138, USA}
\newcommand{\heng}{School of Engineering \& Applied Sciences, Harvard University, Cambridge, MA 02138, USA}
\newcommand{\rice}{Department of Materials Science and NanoEngineering, Rice University, Houston, TX, 77005, USA}
\newcommand{\mitphys}{Department of Physics, Massachusetts Institute of Technology, Cambridge, MA, 02139, USA}
\newcommand{\stanford}{Department of Physics, Stanford University, Stanford, CA 94305, USA}
\newcommand{\wells}{Department of Physics, Wellesley College, Wellesley MA, 02481, USA}

\author{Jiatong Yang}
\affiliation{\wells}
\affiliation{\hphys}
\affiliation{\mitphys}
\author{Benjamin H. November}
\affiliation{\hphys}
\author{Yiting Huang}
\affiliation{\hphys}
\affiliation{\stanford}
\author{S.~Minhal Gardezi}
\affiliation{\wells}
\affiliation{\heng}
\author{Harris Pirie}
\affiliation{\hphys}
\affiliation{\rice}
\author{Jennifer E. Hoffman}
\email{jhoffman@g.harvard.edu}
\affiliation{\hphys}
\affiliation{\heng}

\date{\today}

\begin{abstract}
Flat bands with vanishing group velocity and quenched kinetic energy provide fertile ground for correlated states and exotic phenomena such as unconventional superconductivity. Yet engineering and characterizing flat bands in quantum materials is time-consuming and costly, limiting the exploration of the vast design space of possible flat band systems. Acoustic metamaterials offer an accessible alternative: they can be easily simulated, cheaply fabricated, and quickly measured. Here we present a complete workflow to rapidly prototype flat band systems using acoustic metamaterials. Our design directly implements a tight-binding model using air cavities as lattice sites connected by channels that control hopping, allowing it to generalize to diverse lattice geometries. Using the kagome lattice as a proof-of-concept, we demonstrate excellent agreement between tight-binding theory, finite-element simulations, and experimental measurements. We then design a family of extended kagome lattices whose added sites cancel successively longer-range hopping, further narrowing the flat band. With fabrication and measurement requiring hours rather than months and a total cost orders of magnitude lower than quantum materials, our approach enables rapid iteration through candidate lattices, facilitating the discovery of new flat band physics in quantum materials. 
\end{abstract}

\maketitle

\ptitle{Flat band importance}
Electronic flat bands provide a rich platform for novel quantum phases due to their low kinetic energy and high degeneracy. Their low group velocity emphasizes the role of electron interactions, which would usually be dwarfed by the kinetic energy of dispersive bands. Such strong interactions underpin phenomena such as magnetism and Kondo effects \cite{kouwenhoven_revival_2001}, fractionalization \cite{senthil_Z2-gauge_2000}, and strange metals and hydrodynamic flow \cite{lucas_resistivity_2017}. Degeneracy enables the highly-entangled many-body states required for topological quantum computing \cite{nayak_non-abelian_2008}.
Flat bands can promote superconductivity both by concentrating many states near the Fermi level for pairing and by increasing the relative importance of pairing interactions compared with kinetic energy \cite{bardeen_theory_1957, KopninPRB2011}, while the quantum geometry of the flat band can provide finite superfluid stiffness despite the vanishing conventional band-velocity contribution \cite{julku_geometric_2016}.
Recent experiments show rapid developments in flat band physics \cite{checkelsky_flat_2024}, such as the correlated insulating phase and unconventional superconductivity in magic-angle twisted bilayer graphene \cite{cao_unconventional_2018}, as well as exotic properties in kagome lattices \cite{tang_high-temperature_2011, neupert_fractional_2011, sun_nearly_2011, zhu_interaction-driven_2016, ye_hopping_2024}. The search is now expanding towards materials with even flatter bands and different symmetry representations \cite{chen_emergent_2024}.
However, the synthesis, characterization, and optimization of each candidate quantum material among the vast phase space of possibilities is time-consuming and costly.

\ptitle{Acoustic metamaterials as quantum mimics}
To address these challenges, other flexible platforms have been employed to study flat band systems, including cold atoms trapped in optical lattices \cite{jo_ultracold_2012, taie_coherent_2015, ozawa_interaction-driven_2017, taie_spatial_2020}, photonic metamaterials \cite{nakata_observation_2012, baboux_bosonic_2016, zong_observation_2016}, mechanical metamaterials \cite{rosendo_lopez_flat_2020, karki_non-singular_2023, samak_direct_2024}, and acoustic metamaterials \cite{ding_experimental_2019, wen_acoustic_2019, xiao_geometric_2015, xue_acoustic_2019, qi_acoustic_2020, xiao_synthetic_2015, li_weyl_2018, gardezi_simulating_2021, han_all-angle_2025, zhang_observation_2025, fan_acoustic_2025, cheng_three-dimensional_2024, li_acoustic_2024, mo_observation_2025, jiang_experimental_2021, riva_creating_2025, shi_spin-1_2020, shen_observing_2022, ZhuNatCom2026}. Among these, acoustic metamaterials are especially well suited for a rapid prototyping workflow as they can be simulated in minutes, fabricated at modest cost, and measured in hours. Within the past decade, acoustic metamaterials have reproduced many apparently quantum effects, including quantum Hall phases in graphene \cite{ding_experimental_2019, wen_acoustic_2019}, topological phases with quantized multiple moments \cite{xiao_geometric_2015, xue_acoustic_2019, qi_acoustic_2020}, and topological semimetals \cite{xiao_synthetic_2015, li_weyl_2018}. More recently, acoustic flat bands have been realized and studied in twisted moir\'e lattices \cite{gardezi_simulating_2021, han_all-angle_2025, zhang_observation_2025, fan_acoustic_2025}, Landau level systems \cite{wen_acoustic_2019, cheng_three-dimensional_2024, li_acoustic_2024, mo_observation_2025}, kagome lattices \cite{jiang_experimental_2021, riva_creating_2025}, and others \cite{shi_spin-1_2020, shen_observing_2022, ZhuNatCom2026}. These studies establish the feasibility of creating and measuring acoustic flat bands, and initial workflows that connect simulation, fabrication, and experiment are beginning to emerge.  However, to fully unlock the potential of acoustic platforms as a general-purpose tool for flat-band prototyping, more systematic and broadly accessible design-to-experiment workflows are needed. 

\ptitle{Here we show}
Here, we present a complete workflow for rapidly prototyping flat band lattices with acoustic metamaterials. We demonstrate each step using the kagome lattice as a proof-of-concept. Starting from the tight-binding Hamiltonian, we design an analogous acoustic metamaterial where air cavities act as lattice sites, and connecting channels control the hopping strength between them. We use COMSOL to optimize the design parameters and predict the acoustic band structure. We then CNC-mill the metamaterial from aluminum, though the design principles apply to any other fabrication method and material with sufficient acoustic impedance contrast. Using a simple experimental setup, we measure the acoustic response at each lattice site, and directly visualize both the compact localized states in real space and the complete band structure in momentum space. Our experimental results show good agreement with both tight-binding theory and COMSOL simulations, demonstrating a practical platform for exploring flat band physics.

\begin{figure}[t!]
    \centering
    \includegraphics[clip=true, width=\columnwidth]{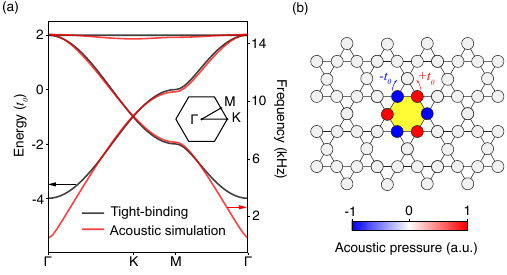}
    \caption{Kagome lattices host a geometric flat band. (a) Tight-binding (black) and COMSOL-computed acoustic (red) band structures of the kagome lattice show good agreement. Both feature two dispersive bands forming a Dirac cone at the $K$ point, and a flat band that touches them at the $\Gamma$ point. (b) The flat band arises from destructive interference between sites of opposite phase (blue and red), which cancels hopping to neighboring sites and confines the mode to a single hexagonal plaquette (yellow).} 
    \label{fig:intro}
\end{figure}

\ptitle{Kagome lattice}
The kagome lattice is a canonical example that hosts a geometrically-protected flat band. In a tight-binding model with only nearest-neighbor hopping $t_0$, it contains two dispersive bands forming Dirac cones at the $K$ and $K'$ points of the Brillouin zone, and a perfectly flat band that touches them at the $\Gamma$ point (Fig.~\ref{fig:intro}(a)). This flat band originates from the destructive interference between hopping terms: its real-space mode has equal amplitude but alternating sign on the six sites of the hexagon, so that hopping to neighboring sites interferes destructively, and the wave function remains confined to a single hexagonal plaquette (Fig.~\ref{fig:intro}(b)). The result is a compact localized state with zero group velocity---the defining characteristic of a perfectly flat band. In a more realistic model, next-nearest-neighbor hopping $t_1$ introduces a slight dispersion, but the group velocity remains near zero. 

\ptitle{Metamaterial design}
We designed an acoustic metamaterial that directly implements the kagome tight-binding model. In our system, an array of air cavities hold acoustic modes analogous to atomic orbitals, while interconnecting air channels control $t_0$. We focus on the lowest-frequency cavity modes, which exhibit nearly uniform pressure and are analogous to atomic $s$ orbitals. The cavities and channels are cut into a solid lattice layer and sealed within boundary sheets to contain the acoustic pressure. Because the acoustic impedance of the solid exceeds that of air by more than two orders of magnitude, sound waves propagate almost entirely through the engineered air lattice, with negligible leakage into the surrounding solid. This impedance confinement ensures that the acoustic band structure is determined primarily by the tight-binding geometry, rather than material choice. Our design allows straightforward implementation and tuning of diverse lattice geometries using either subtractive (cutting, milling) or additive (3D printing) fabrication techniques. 

\begin{figure}[t]
    \centering
    \includegraphics[clip=true, width=\columnwidth]{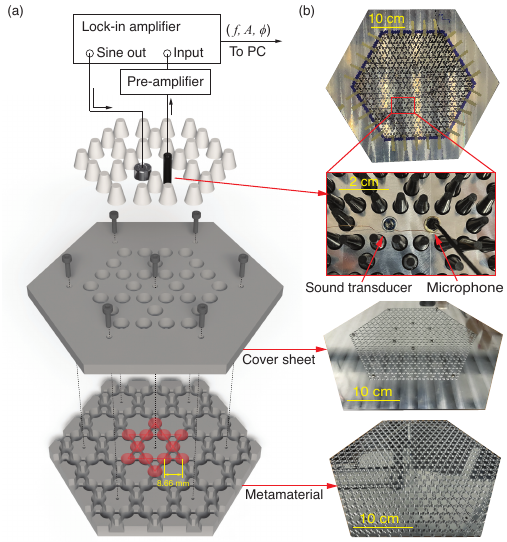}
    \caption{Experimental apparatus to measure acoustic flat bands (a) A lock-in amplifier generates a sinusoidal drive signal for the sound transducer and performs phase-sensitive detection of the microphone signal via a pre-amplifier. The metamaterial consists of a total of 1296 cavities arranged in a kagome lattice (highlighted in red) enclosed by a removable cover sheet. (b) Photographs of the fabricated acoustic kagome metamaterial.}
    \label{fig:setup}
\end{figure}

\ptitle{COMSOL simulation}
We quantified our design using finite-element simulations in COMSOL Multiphysics{\textsuperscript{\textregistered}} with the ``Pressure Acoustics, Frequency Domain'' interface from the Acoustics Module. The kagome metamaterial is a 5-mm-thick aluminum layer with air cavities of radius $R=3.5$ mm, channel width $w=2$ mm, and nearest-neighbor distance $a=8.66$ mm. This layer is sandwiched between a 5-mm-thick silicone cover sheet and a 3-mm-thick aluminum base layer. We impedance-match the upper surface of the cover sheet to air and impose a sound-hard boundary condition at the lower surface of the base layer. We use material properties for aluminum (mass density $\rho = 2700$ kg/m$^3$, speed of sound $v = 6300$ m/s), air ($\rho = 1.2$ kg/m$^3$, $v = 343$ m/s), and silicone ($\rho = 1000$ kg/m$^3$, $v = 1050$ m/s). To calculate the acoustic band structure, we simulate a single unit cell with Floquet-periodic boundary conditions on its three pairs of lateral faces.
We use COMSOL's ``Eigenfrequency'' study while sweeping the Bloch wavevector along a high-symmetry path in reciprocal space.
A band structure calculation with 48 $k$-points required only 85 seconds, demonstrating the computational efficiency of the acoustic model. The resulting dispersion reproduces the Dirac cone and flat band of the tight-binding model, despite the longer-range couplings included in the full finite-element geometry (see comparison in Fig.~\ref{fig:intro}(a)). Using the dispersive band opening at the M point yields an estimation of $t_0\approx1.75$ kHz. 

\begin{figure}[t!]
    \centering \includegraphics[clip=true,width=\columnwidth]{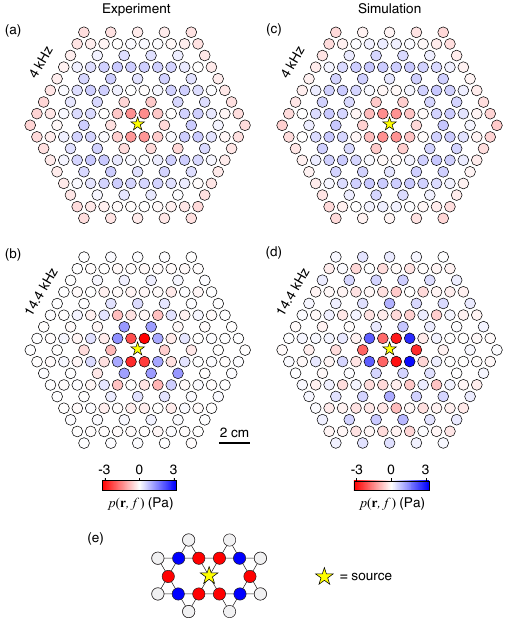}
    \caption{Real-space mapping of dispersive and localized modes in the acoustic kagome lattice. (a)-(b) Experimental measurements of the acoustic field at 4 kHz (dispersive mode) and 14.4 kHz (flat-band mode). The star marks the position of the sound transducer. At 4 kHz, the excitation propagates across the lattice, while at 14.4 kHz the response is strongly confined around the source, demonstrating flat-band localization. (c)-(d) COMSOL simulations of the corresponding modes show excellent agreement with experiment. We emphasize that the identical quantitative color scale for pressure is used for all measurements and calculations shown here in (a)-(d). (e) Schematic of the compact localized state expected from a single central source.} 
    \label{fig:exp_real_space}
\end{figure}

\begin{figure}[htb!]
    \centering
    \includegraphics[clip=true,width=\columnwidth]{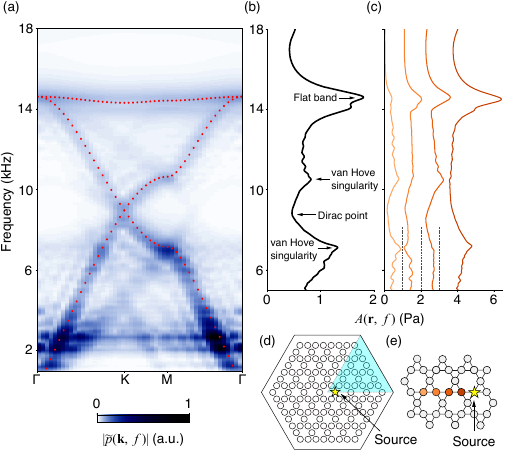}
    \caption{Spectroscopic evidence of a localized flat band in an acoustic kagome metamaterial. (a) The experimentally reconstructed band structure (blue colorscale) obtained from the spatial Fourier transform of the measured pressure fields matches the simulated band structure (red points). There are several less prominent bands that arise from the boundary effects in the finite-size Fourier transform.
    The inset shows a schematic of the Brillouin zone sweep. (b) The averaged sound-pressure spectrum from the 14 cavities nearest the source exhibits the characteristic features of a kagome lattice: a dip at the Dirac point, peaks at the two Van Hove singularities, and a pronounced resonance at the flat-band frequency. (c) Individual cavity spectra show that the flat-band resonance decays rapidly with distance from the source, while the dispersive modes maintain similar amplitudes across the lattice, confirming the localized nature of the flat-band state. The spectra have shifted origins (marked with dashed lines) along the horizontal axis for clarity. (d) Schematic of the kagome metamaterial (not drawn to scale and doesn't include all cavities). The cyan shading shows the cavities where the band structure data in (a) were obtained. (e) Color-coded schematic of the cavities in which the response curves in (c) were acquired.}
    \label{fig:exp_band}
\end{figure}

% K-space resolution: 
% SI units: 0.36 cm^-1 in the gamma-K direction. 0.47 cm^-1 in gamma-M direction.
% Percent of Brillouin zone: 7.3% in the gamma-K direction and 9.7% in the gamma-M direction.

\begin{figure*}[ht!]
    \centering
    \includegraphics[clip=true,width=\textwidth]{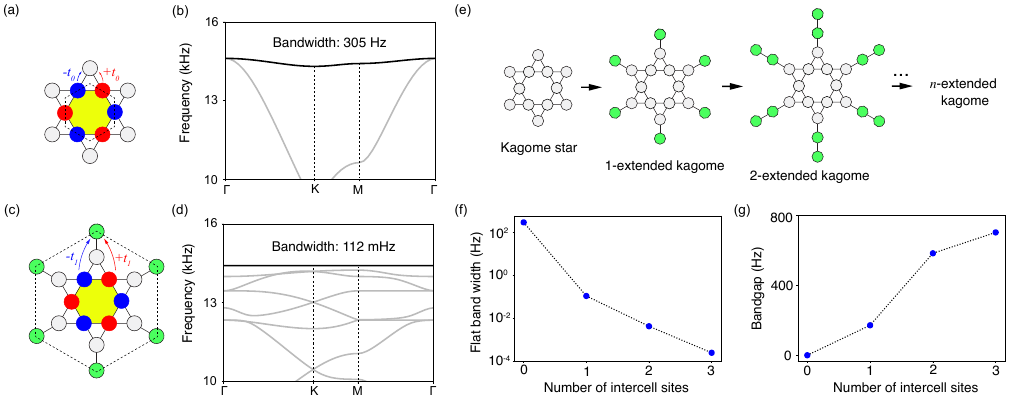}
    \caption{The extended kagome lattice hosts an isolated, ultra-flat band. (a) Unit cell of the kagome lattice and corresponding (b) COMSOL simulation showing flat band width of 305 Hz. (c) Unit cell of extended kagome. The green sites cancel the next-nearest-neighbor hopping $t_1$ from the red and blue sites, further localizing the mode to the yellow region. (d) COMSOL-simulated band structure of extended kagome. The top band is unpinned from the rest of the dispersive bands and is flatter than the kagome flat band, with a bandwidth of 112 mHz. This simulation was done with the same setup and geometry as the kagome simulation, with an aluminum metamaterial and silicone rubber cover sheet. (e) The unit cell of $n$-extended kagome is constructed by adding $2n$ cavities (green) to the kagome star. (f) The bandwidth of the flat band decreases almost exponentially with the number of intercell sites. (g) The band gap between the flat and dispersive bands increases with the number of intercell sites added.}
    \label{fig:extended_kagome}
\end{figure*}

% Simulated flat band quality factors: Kagome has $Q\sim 47$ (bandwidth / band frequency). Extended kagome has $Q\sim 128,798$. 

\ptitle{Experimental setup}
We designed an experimental apparatus for accessible, efficient, and precise acoustic mapping in both real and momentum space (Fig.~\ref{fig:setup}). The kagome lattice and base layer were CNC-machined from a single piece of aluminum and sealed with a removable aluminum cover plate. The cover contains an array of small access holes above each cavity, allowing us to excite acoustic modes with a sound transducer in one cavity and measure the response by scanning a microphone across all other sites. Unused holes are sealed with rubber plugs, creating a condition similar to the simulated silicone cover sheet. We use a lock-in amplifier both to generate a monofrequency transducer tone and to enable phase-sensitive detection at each cavity location $\mathbf{r_i}$: the microphone output is amplified then returned to the lock-in to extract the amplitude $A(\mathbf{r_i}, f)$ and phase $\phi(\mathbf{r_i}, f)$ at the driving frequency $f$. The steady-state acoustic wave function can be reconstructed by calculating the time-dependent pressure 
\begin{equation}
    p(\mathbf{r_i}, f, t) = A(\mathbf{r_i}, f)e^{-i(2\pi ft+\phi(\mathbf{r_i}, f))}.
\end{equation}
\noindent Typically, we focus on the time-independent solutions 
\begin{equation}
    p(\mathbf{r_i}, f) = A(\mathbf{r_i}, f)e^{-i\phi(\mathbf{r_i}, f)}.
\end{equation}
\noindent We compute the complete two-dimensional band structure from the Fourier transform of $p(\mathbf{r_i}, f)$
\begin{equation}
    \tilde{p}_n (\mathbf{k}, f) = \sum_i A(\mathbf{r_i}, f)e^{-i\phi(\mathbf{r_i}, f)} e^{-i\mathbf{k}\cdot \mathbf{r_i}}, 
\end{equation}
\noindent where $i$ is summing over all cavities in the sublattice $n$. Then we sum over all three kagome sublattices to obtain the average Fourier amplitude 
\begin{equation} \label{eq:fourier_sum}
    |\tilde{p} (\mathbf{k}, f)| = \frac{1}{3} \sum_n |\tilde{p}_n (\mathbf{k}, f)|.
\end{equation}
Measuring a single cavity with high signal-to-noise takes approximately 0.5 seconds per frequency point. Including manual microphone repositioning, measuring a 100--200 cavity lattice over its entire bandwidth requires 4--8 hours. 

\ptitle{Experimental data - real space}
Our apparatus can directly image real-space acoustic wave functions, making it straightforward to distinguish between the propagating modes of dispersive bands and the localized states of a flat band. We first excited the lattice at a frequency of 4 kHz, within the lower dispersive band. This excitation induced a propagating wave spreading from the transducer to fill the entire lattice (Fig.~\ref{fig:exp_real_space}(a) and supplementary video in Appendix \ref{app:movie}). In contrast, at the flat-band frequency of 14.4 kHz, the response remained tightly confined near the source and decayed rapidly beyond the first unit cell (Fig.~\ref{fig:exp_real_space}(b)). This rapid decay is a consequence of destructive interference in the kagome lattice, which we directly observed through the alternating phases between neighboring sites. The overall response can be understood as a superposition of two compact localized states of the form shown in Fig.~\ref{fig:intro}(a), as highlighted schematically in Fig.~\ref{fig:exp_real_space}(e). The localized flat-band state reached a higher peak pressure amplitude than the dispersive bands---a consequence of concentrating the acoustic energy in a confined region. Our experimental measurements are in agreement with the finite-element simulations for both the propagating mode (Fig.~\ref{fig:exp_real_space}(c)) and the confined flat-band mode (Fig.~\ref{fig:exp_real_space}(d)). We used COMSOL's ``Frequency Domain'' study to simulate the acoustic response of the entire metamaterial driven by a point source placed inside a cavity near the middle.

\ptitle{Experimental data - band structure}
To complete our characterization, we reconstructed the full acoustic band structure in Fig.~\ref{fig:exp_band} by calculating the average Fourier amplitude shown in Eq.~\eqref{eq:fourier_sum}. To achieve high $k$-space resolution of around 0.4 $\text{cm}^{-1}$ within a reasonable measurement time, we sampled a triangular region containing one-sixth of the lattice sites (Fig.~\ref{fig:exp_band}(d)). Then we symmetrized the Fourier-transformed data from Eq.~\eqref{eq:fourier_sum} to recover the sixfold rotational symmetry of the Brillouin zone. In total, we measured 185 cavities in around 8 hours, though this process could be substantially accelerated using a multiplexed or automated scanning system. The resulting experimental band structure agrees with both the finite-element simulations and tight-binding calculations, displaying a clear Dirac cone at the $K$ point and a high-intensity flat band at around 14.4 kHz (Fig.~\ref{fig:exp_band}(a)). Averaging the amplitude over the 14 cavities nearest to the source provides an estimate of the local density of states, revealing the characteristic kagome-lattice features: a dip at the Dirac point, two van Hove peaks, and a dominant flat band maximum (Fig.~\ref{fig:exp_band}(b)). Single-site spectra recorded at increasing distances from the source confirm the localized nature of the flat band, as its spectral peak decays much faster than those of the propagating modes (Fig.~\ref{fig:exp_band}(c)). 

\ptitle{Introduce extended kagome and simulation}
Our rapid prototyping workflow can be used not only to study existing flat band systems, but also to explore lattices that do not exist naturally.  As a demonstration, we designed a family of extended kagome lattices whose flat bands become progressively flatter. In the conventional kagome lattice, next-nearest neighbor hopping $t_1$ prevents a perfectly flat band (Fig.~\ref{fig:extended_kagome}(a,b)). Our extended kagome lattice adds one site outside each kagome star, so that both $t_0$ and $t_1$ interfere destructively (Fig.~\ref{fig:extended_kagome}(c)). The extended kagome flat band is more than two orders of magnitude flatter than the kagome flat band, and is also unpinned from the rest of the dispersive bands (Fig.~\ref{fig:extended_kagome}(d)). Adding $n$ sites in between the kagome stars cancels the hopping terms up to $t_n$, creating a series of ``$n$-extended kagome'' lattices with progressively flatter bands (Fig.~\ref{fig:extended_kagome}(e)). We simulated the first four lattices in the series with COMSOL (Fig.~\ref{fig:extended_kagome}(f,g)). The flat band narrows nearly exponentially with increased $n$, while its gap to the dispersive bands widens. Each added intercell site pushes neighboring kagome units further apart, exponentially suppressing the hopping that sets the residual bandwidth.

\ptitle{Conclusion}
Our work establishes a practical bridge between theoretical flat-band models and experimental realization, transforming studies from months-long materials synthesis to hours-long table-top experiments. By directly imaging both the real-space acoustic wave functions and the full momentum-space acoustic band structure, we simultaneously captured the compact localized states and near-zero group velocity of the kagome lattice---features that underpin correlation, topology, and superconductivity in electronic systems. Beyond the simple kagome lattice, our framework offers a route to experimentally explore the vast phase space of flat-band systems, including large moir\'{e} geometries where conventional theoretical and numerical approaches reach their limits \cite{carr_electronic-structure_2020} and artificially designed lattices that do not yet exist in nature such as the ``$n$-extended kagome'' lattice.

%\bibliography{refs}

\newpage
\onecolumngrid

\appendix
\section{Time evolution} \label{app:movie}
\vspace{-3mm}
Our apparatus can directly image real-space acoustic wave functions. In addition to the snapshots presented in Fig.~4 of the main text, we include a video showing the time evolution of excitations within the dispersive regime (4 kHz) and the flat band regime (14.4 kHz). At 4 kHz, the transducer induced a propagating wave spreading from the center to fill the entire lattice, whereas at the flat-band frequency of 14.4 kHz, the response remained tightly confined near the source and decayed rapidly beyond the first unit cell. Measured acoustic pressure and COMSOL simulation show excellent agreement in both frequency regimes.\\
\url{https://hoffman.physics.harvard.edu/publications/Yang-metamaterial-kagome-flat-band.gif}

\vspace{-3mm}
\section{Cost table} \label{app:cost_table}
\vspace{-3mm}
The total equipment cost for our setup was about \$5,700 in 2025, with the lock-in amplifier representing the bulk of the expense (see Table \ref{tab:material_cost} for detailed cost breakdown). These equipments can be reused for measuring other metamaterials, so the cost per experiment can be reduced. The aluminum kagome lattice cost approximately \$1,750 to fabricate. This could be reduced with 3D printing at the expense of lower acoustic impedance and reduced quality factor, which may still be acceptable for some applications~\cite{wen_acoustic_2019, zhang_observation_2025}. Overall, the total cost of our approach is orders of magnitude lower than synthesizing and characterizing quantum materials, which can easily reach tens of thousands of dollars when considering equipment, materials, and labor.

\begin{center}
\begin{table}[h!]
\small
\renewcommand{\arraystretch}{1.2} 
\setlength{\tabcolsep}{8pt}       
\begin{tabular}{l l r r r}
\textbf{Product} & \textbf{Part No. / Description} & \textbf{Qty.} & \textbf{Unit Price (USD)} & \textbf{Total (USD)} \\
\hline
Transducer & Knowles SR6438NWS-000 & 1 & 7.66 & 7.66 \\
Microphone & DPA 6060-OC-U-B00 & 1 & 640.00 & 640.00 \\
Microlock to XLR adapter & DPA DAD9001 & 1 & 125.00 & 125.00 \\
Microphone preamplifier & M-Audio DMP3 & 1 & 100.00 & 100.00 \\
Lock-in amplifier & Stanford Research SR830 & 1 & 4,795.00 & 4,795.00 \\
\hline
\multicolumn{4}{r}{\textbf{Measurement setup total:}} & \textbf{5,667.66} \\
\\
Silicone plugs & STP102 & 1,300 & 0.0417 & 54.21 \\
Lattice layer (CNC machined) & Aluminum & 1 & 1,005.52 & 1,005.52 \\
Cover sheet (CNC machined) & Aluminum & 1 & 676.67 & 676.67 \\
\hline
\multicolumn{4}{r}{\textbf{Metamaterial total:}} & \textbf{1,736.40} \\
\end{tabular}
\caption{Breakdown of material and setup costs. The measurement setup includes reusable instruments, dominated by the cost of the lock-in amplifier. The metamaterial costs include components specific to prototyping a single lattice.}
\label{tab:material_cost}
\end{table}
\end{center}

\end{document}